\documentclass[epj,final]{svjour}
\usepackage{psfig}

\begin{document}
\title{Transport analysis of in-medium hadron effects in $pA$ and
$AA$
collisions}

\author{W. Cassing
%%\inst{1}
%%%\thanks is optional - remove next line if not needed
\thanks{In collaboration with E. L. Bratkovskaya,
M. B\"uscher, Ye. S. Golubeva, V. Grishina, V. Hejny, V. Metag, J.
Messchendorp, B. Kamys, L. A. Kondratyuk, P. Kulessa, Z. Rudy, S.
Schadmand, A. A. Sibirtsev, H. Str\"oher}
%%%\emph{Present address:} Insert the address here if needed}%
}                     % Do not remove
\institute{
Institut f\"ur Theoretische Physik, Justus Liebig Universit\"at
Giessen, D-35392 Giessen, Germany}
\date{Received: date / Revised version: date}
% The correct dates will be entered by Springer
\abstract{ The production and decay of vector mesons
($\rho, \omega$) in $pA$ and $AA$ reactions  is studied
with particular emphasis on their in-medium spectral functions. It
is explored within transport calculations if hadronic in-medium
decays like $\pi^+\pi^-$ or $\pi^0 \gamma$ might provide
complementary information to their dilepton ($e^+e^-$) decays.
Whereas the $\pi^+ \pi^-$ signal from the $\rho$-meson  is found
to be strongly distorted by pion rescattering, the $\omega$-meson
Dalitz decay to $\pi^0 \gamma$ appears promising even for more
heavy nuclei in $\gamma A$ and $pA$ reactions.
Furthermore, the influence of nucleon and
kaon/antikaon potentials on the $K^\pm$ yields and spectra in $pA$ collisions is
calculated and compared to the recent data from the ANKE
Collaboration.
 % end of PACS codes
} %end of abstract

\maketitle

\section{Introduction}
The modification of the meson properties
\cite{Medium5,Medium6} in nuclear matter has become a challenging
subject in hadron physics from $\gamma A$, $p A$ and $A A$
collisions. Here the dilepton ($e^+e^-$) radiation from $\rho$'s
and $\omega$'s propagating in finite density nuclear matter is
directly proportional to their spectral function which becomes
distorted in the medium due to the interactions with nucleons.
Apart from the vacuum width $\Gamma^0_V \ (V= \rho, \omega$) these
modifications are described by the real and imaginary part of the
retarded  self energies $\Sigma_V$, where the real part $\Re
\Sigma_V$ yields a shift of the meson mass pole and the additional
imaginary part $\Im \Sigma_V$ (half) the collisional broadening of
the vector meson in the medium. We recall that the meson self
energy in the $t-\rho$ approximation is proportional to the
complex forward $V N$ scattering amplitude $f_{VN}(P,0)$ and the
nuclear density $\rho(X)$, i.e. $ \Sigma_V(P,X) = - 4 \pi \rho(X)
f_{V N}(P,0)$. The scattering amplitude itself, furthermore, obeys
dispersion relations between the real and imaginary parts
\cite{d3} while the imaginary part can be determined from the
total $V N$ cross section according to the optical theorem. Thus
the vector meson spectral function,
\begin{eqnarray}
A_V(X,\!P){\sim} \frac{\Im \Sigma_V (X,\!P)}{(P^2{-}
M_{V}^{2}{-}\Re\Sigma_V(X,\!P))^{2}{+} \Im \Sigma_V (X,\!P)^{2}} ,
 \label{spectral}
\end{eqnarray} can be constructed
once the $V N$ elastic and inelastic cross sections are known.
Note that in (\ref{spectral}) all quantities depend on space-time
$X$ and 4-momentum $P$.

\section{Production and decay of vector mesons}
Though strong modifications of the $\rho$-meson properties in
relativistic $A+A$ reactions have been reported by the CERES
Collaboration \cite{CERES}, the interpretation of the data is not
unique since the latter may be explained by a dropping of the pole
mass as well as by ordinary nuclear many-body effects
\cite{Medium5,Medium6}. In general, the medium effects on the
$\rho$-meson are expected to increase when decreasing the
bombarding energy from 160 A$\cdot$GeV down to about 2 A$\cdot$GeV
for central $Au+Au$ collisions \cite{excita}. At the lower range
of bombarding energies we expect new data coming up from the HADES
Collaboration in the near future \cite{HADES}. Nevertheless, it is
inevitable to study also $\gamma  A$ and $p A$ reactions where the
time evolution of the baryon density is much better under control
and the vector-meson spectral function is essentially tested at
density $\rho_0$ in heavy nuclei \cite{m3,Bratnew}.

The in-medium vector meson spectral functions
can be measured directly by the leptonic decay $V {\to}e^+e^-$
(cf. Refs. \cite{m3,Bratnew}), the strong decay $\rho^0
\rightarrow \pi^+ \pi^-$ or the Dalitz decay
$\omega{\to}\pi^0\gamma$, respectively. Thus it remains to be seen
which of the decay modes is most effective for experimental
studies.

The actual transport calculations have been performed for $p A$
and $\gamma  A$ collisions by introducing a real and imaginary
part of the vector meson self energy and width as
\begin{equation}
\label{poten}
 U_V = \frac{\Re \Sigma_V}{2 M_0} \simeq M_0 \
\beta \frac{ \rho(X)}{\rho_0}, \end{equation}
\begin{equation}  \Gamma^\ast =
\frac{\Im \Sigma_V}{2 M_0} \simeq \Gamma_V^0{+}\Gamma_{coll}
\frac{\rho(X)}{\rho_0},
\end{equation}
in the $t{-}\rho$ approximation. Here $M_0$ and $\Gamma_V^0$
denote the bare mass and width of the vector meson in vacuum while
$\rho(X)$ is the local baryon density ($\rho_0$=0.16~fm$^{-3}$).
The parameter $\beta{\simeq}$--0.16 was adopted from the models
discussed in Refs.\cite{Medium5,Medium6}. The predictions for the
$\omega$-meson collisional width $\Gamma_{coll}$ at density
$\rho_0$ range from 20 to 50~MeV \cite{OmegaN}-- depending on the
number of $\omega{N}$ final channels taken into account --  while
the collisional width of the $\rho$-meson should be about 100 --
120 MeV at $\rho_0$ due to the strong coupling to baryon
resonances (cf. Fig. 6 of Ref.\cite{d3}).

As a first remark we quote the result from Ref. \cite{Model}, where it
has been found that the $\pi^+ \pi^-$ decay mode of the $\rho^0$
in nuclei is not well suited to reconstruct the in-medium $\rho^0$
spectral function except for very light nuclei due to the strong
$\pi^\pm$ final state interactions.
\begin{figure}[t]
\centerline{\psfig{figure=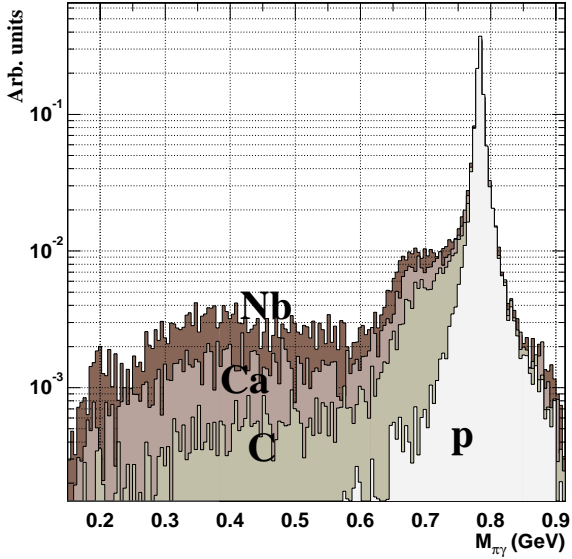,width=7.5cm}} \caption{The
$\pi^0 \gamma$ invariant mass distribution from $\omega$-meson
decays in $\gamma A$ reactions including an attractive mass shift
and collisional broadening at $E_\gamma =$ 1.2 GeV for a proton,
$^{12}C, \ ^{40}Ca$ and $Nb$ target according to Ref.
\protect\cite{Johan01}. The spectra have been normalized to the
vacuum decay peak at 0.783 GeV.}
 \label{bild1}
\end{figure}
The situation changes for the in-medium $\omega$-meson Dalitz
decay since here only a single pion might rescatter whereas the
photon escapes practically without reinteraction. In this case it
is found \cite{Hejny,Johan01} that though most of the
$\omega$-mesons -- produced in photon induced reactions on nuclear
targets -- decay in the vacuum, there is a sizeable contribution
of $\omega$'s decaying in the medium leading to $\pi^0 \gamma$
pairs of invariant mass 0.6 -- 0.9 GeV, however, also $\pi^0$
rescattering gives a substantial background essentially below 0.65
GeV.

As an example we show in Fig. 1 the $\pi^0 \gamma$ invariant mass
distribution from $\omega$-meson decays including an attractive
mass shift and collisional broadening at $E_\gamma =$ 1.2 GeV for
a proton, $^{12}C,\  ^{40}Ca$ and $Nb$ target according to Ref.
\cite{Johan01}. The background -- in the invariant mass range 0.6
$\leq M \leq$ 0.9 GeV of interest -- can be suppressed effectively
by kinematical cuts on higher $\pi^0$ energies \cite{Johan01} or
angular correlations  between the photon and the pion
\cite{Golub01} which provides good perspectives for the $\pi^0
\gamma$ decay mode of the $\omega$-meson.
\begin{figure}[b]
\centerline{\psfig{figure=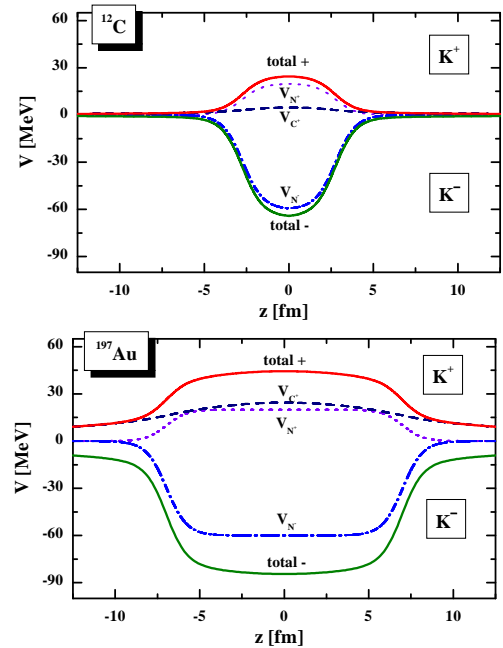,width=6.5cm}}
 \caption{The effective nuclear (\ref{par}) (dotted lines) and Coulomb potentials
(dashed lines) for $K^+$ and $K^-$ mesons
 in case of a $^{12}C$ and $^{197}Au$ target (see text).}
 \label{bild2}
\end{figure}

\section{$K^\pm$ production}

The in-medium properties of kaons (and antikaons) show up in their
suppressed (enhanced) yields in $p A$ or $A A$ collisions
\cite{Ref-6,Cass96} as well as in their low momentum spectra
\cite{Medium5}, which are most sensitive to the meson nuclear potentials.
The experiments of the FOPI \cite{FOPI} and KaoS Collaboration
\cite{kaos} on nucleus-nucleus collisions at SIS energies
have demonstrated that the abundancy and collective
flow of kaons and antikaons indicate slightly repulsive potentials
for $K^+$ and more strongly attractive potentials for $K^-$.
However, as pointed out in Ref. \cite{Cass96}, the $K^-$ yield is
dominated by the strange flavor exchange reaction $\pi + Y
\leftrightarrow \bar{K} N$ which by strangeness conservation strongly
links the $K^+$ and $K^-$ yields. The problem is that the cross
section for the latter reaction at high baryon density is not
sufficiently known and that data from $A A$ reactions do not allow for a unique
conclusion. Thus $K^\pm$ production in $p A$ reactions is a
necessary and complementary study since in these type of reactions the
$\pi + Y \rightarrow \bar{K} N$ channel plays a minor role and the
$K^\pm$ potentials can be extracted with less ambiguities.

In $p A$ reactions the $K^\pm$ mesons are dominantly produced with
a finite momentum relative to the nucleus at rest and -- once they
do not rescatter elastically or inelastically -- are accelerated
or decelerated by the nuclear and Coulomb potential. For
orientation the effective nuclear and Coulomb potentials are
displayed in Fig. 2 for a $^{12}C$ and $^{197}Au$ target as a
function of the coordinate $z$. Here the $K^\pm$ nuclear potentials have
been approximated by
\begin{equation} \label{par}
V_N^\pm({\bf r}) = V_0^\pm \frac{\rho({\bf r})}{\rho_0}
\end{equation}
with $V_0^+$ = 20 MeV \cite{sibirtsev} and $V_0^-$ = - 60 MeV
\cite{Ramos}. As seen from Fig. 2 the Coulomb potential $V_C$ plays only
a minor role for $^{12}C$, however, is even larger than the $K^+$
potential in case of a $^{197}Au$ target. Thus the minimum kinetic
energy of a $K^+$ meson -- produced in the center of a $Au$-target
-- (without rescattering) is about 45 MeV in the continuum and
about 23 MeV in case of $^{12}C$. Kaons produced at the nuclear
surface will have a minimum kinetic energy that is determined by
 $V_C$ at the point of production. Thus $K^+$ ratios from heavy to
 light targets have to show a strong suppression for low $K^+$
 momenta. In fact, it has been
demonstrated in Ref. \cite{Ref-7} that ratios of cross sections for light and heavy
targets in $p A$ reactions provide a sensitive tool to measure the
$K^+$ potentials in an almost model independent way.

 The situation is quite different for antikaons since
 $K^-$ produced with low kinetic energy in the medium cannot
 escape to the continuum without elastic rescattering. Most of
 them are absorbed in the flavor exchange reaction $K^- N
 \rightarrow Y+\pi$ or, if their energy $E_K <$ 0, they might
 form antikaonic atoms when escaping the nuclear medium.

\begin{figure}[t]
\centerline{\psfig{figure=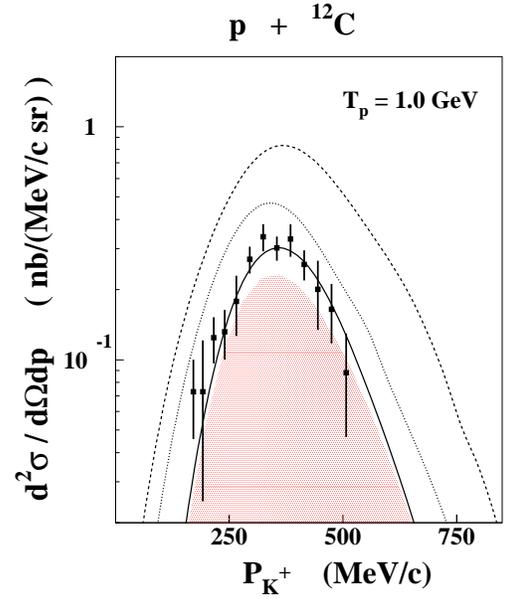,width=6.5cm}}
 \caption{The
calculated differential $K^+$ spectra for $p+C$ at 1.0 GeV for
$\theta_{lab} \leq 12^0$  within the acceptance of the ANKE
spectrometer (from Ref. \protect\cite{Ref-6}) in comparison to the
data from \cite{ANKE}. The dotted (middle) line is obtained from
transport calculations without baryon and kaon potentials, the
dashed (top) line shows the results with baryon potentials
included while the solid line corresponds to calculations
including both, nucleon and kaon potentials. }
 \label{bild8}
\end{figure}

We now turn to the kinematical conditions of the ANKE experiments
at COSY-J\"ulich, that have taken  $K^+$ spectra in forward
direction for $\theta_{lab} \leq 12^o$.  The calculated
differential $K^+$ spectra for $p+^{12}C$ at 1.0 GeV for
$\theta_{lab} \leq 12^0$  are displayed in Fig. \ref{bild8} in
comparison to the data from Ref. \cite{ANKE}. The dotted line is
obtained from transport calculations without baryon and kaon
potentials, the dashed line shows the result with baryon
potentials included while the solid line corresponds to a
calculation with both, nucleon and kaon potentials. At this low
bombarding energy the net attractive baryon potential in the final
state enhances the $K^+$ yield by about a factor of 2 whereas the
additional repulsive $K^+$ potential leads to a decrease by a
factor $\sim$ 3. The data from Ref. \cite{ANKE} are rather well
described by the calculations that include the baryon and $K^+$
potentials (solid line), whereas the other limits clearly fail.
This might be considered as a first indication for the observation
of a repulsive $K^+$ potential in $p A$ reactions, however, a full
systematics in target mass $A$ and laboratory energy $T_{lab}$
will be needed to pin down this effect unambiguously.

The shaded area in Fig. 3 indicates the contributions from the
two-step mechanisms $\Delta N \rightarrow K^+ Y N$ and $\pi N
\rightarrow K^+ Y$, respectively, for the case of nucleon and kaon
potentials included (solid line). Thus  secondary ($\Delta$ and
$\pi$ induced) reaction channels give the dominant fraction of the
$K^+$ yield at 1 GeV even for the light target $^{12}C$.

\section{Summary}
The production and decay of vector mesons
($\rho, \omega$) especially in $\gamma A$ and $pA$  reactions  has been
 studied with particular emphasis on signals from their in-medium spectral functions.
Whereas the $\pi^+ \pi^-$ decay mode from the $\rho$-meson  is
found to be strongly distorted by pion rescattering with the
surrounding nucleons, the $\omega$-meson Dalitz decay to $\pi^0
\gamma$ appears promising even for more heavy nuclei. Furthermore,
the influence of nucleon and kaon/antikaon potentials on the
$K^\pm$ yields and spectra has been evaluated within the transport
approach. A comparison to the recent data from the ANKE
Collaboration \cite{ANKE} yields a repulsive $K^+$ potential of
$\approx$ 20 MeV at normal nuclear matter density
\cite{Ref-6,Ref-7} which is in line with the data from $AA$
reactions at SIS energies \cite{FOPI,kaos}.

\end{document}